\documentclass[aps,twocolumn,amsmath,amssymb,superscriptaddress]{revtex4-1}
\usepackage{graphicx}
\usepackage{xspace}
\usepackage{bm}
\usepackage[T2A]{fontenc}	 
\usepackage[utf8]{inputenc}	
\usepackage[english]{babel}
\usepackage[unicode,colorlinks=true,citecolor=green,linkcolor=blue]{hyperref}
\usepackage{color}

\def\wse2{WSe$_2$\xspace}
\def\mos2{MoS$_2$\xspace}

\begin{document}
\title{Efficient Electron Spin Relaxation by Chiral Phonons in \wse2 Monolayers}

\author{D. Lagarde}
\affiliation{Universit\'e de Toulouse, INSA-CNRS-UPS, LPCNO, 135 Av. Rangueil, 31077 Toulouse, France}
\author{M. Glazov}
\affiliation{Ioffe Institute, 26 Polytechnicheskaya, 194021 Saint Petersburg, Russia
}
\author{V. Jindal}
\affiliation{Universit\'e de Toulouse, INSA-CNRS-UPS, LPCNO, 135 Av. Rangueil, 31077 Toulouse, France}
\author{K. Mourzidis}
\affiliation{Universit\'e de Toulouse, INSA-CNRS-UPS, LPCNO, 135 Av. Rangueil, 31077 Toulouse, France}
\author{Iann Gerber}
\affiliation{Universit\'e de Toulouse, INSA-CNRS-UPS, LPCNO, 135 Av. Rangueil, 31077 Toulouse, France}
\author{A. Balocchi}
\affiliation{Universit\'e de Toulouse, INSA-CNRS-UPS, LPCNO, 135 Av. Rangueil, 31077 Toulouse, France}
\author{L.~Lombez}
\affiliation{Universit\'e de Toulouse, INSA-CNRS-UPS, LPCNO, 135 Av. Rangueil, 31077 Toulouse, France}
\author{ P. Renucci}
\affiliation{Universit\'e de Toulouse, INSA-CNRS-UPS, LPCNO, 135 Av. Rangueil, 31077 Toulouse, France}
\author{T. Taniguchi}
\affiliation{International Center for Materials Nanoarchitectonics, National Institute for Materials Science, 1-1 Namiki, Tsukuba 305-00044, Japan}
\author{K. Watanabe}
\affiliation{Research Center for Functional Materials, National Institute for Materials Science, 1-1 Namiki, Tsukuba 305-00044, Japan}
\author{C. Robert}
\affiliation{Universit\'e de Toulouse, INSA-CNRS-UPS, LPCNO, 135 Av. Rangueil, 31077 Toulouse, France}
\author{X. Marie}
\email{marie@insa-toulouse.fr}
\affiliation{Universit\'e de Toulouse, INSA-CNRS-UPS, LPCNO, 135 Av. Rangueil, 31077 Toulouse, France}
\affiliation{Institut Universitaire de France, 75231 Paris, France} 

\begin{abstract}
In transition metal dichalcogenide semiconductor monolayers the spin dynamics of electrons is controlled by the original spin-valley locking effect resulting from the interplay between spin-orbit interaction and inversion asymmetry.  As a consequence, for electrons occupying bottom conduction bands, a carrier spin flip occurs only if there is a simultaneous change of valley. However, very little is known about the intra-valley spin relaxation processes. In this work we have performed stationary and time-resolved photoluminescence measurements in high quality \wse2 monolayers. Our experiments highlight an efficient relaxation from bright to dark excitons, due to a fast intra-valley electron transfer from the top to the bottom conduction band with opposite spins. A combination of experiments and theoretical analysis allows us to infer a spin relaxation time of about $\tau_s\sim10~$ps, driven by the interplay between $\Gamma$-valley chiral phonons and spin-orbit mixing.

\end{abstract}

\maketitle

\section{Introduction}\label{sec:Intro}

Two-dimensional semiconductors based on transition metal dichalcogenides (TMD) are characterized by  original optical selection rules which have given rise to numerous theoretical and experimental works, see Refs.~\cite{PRL12dx, NRM16js, RMP18gw} and references therein.
The inversion symmetry breaking together with the spin-orbit interaction leads to a unique coupling of the charge carrier spin and the $\bm k$-space valley physics. The circular polarization of the absorbed or emitted photon ($\sigma^+$ or $\sigma^-$ ) can be directly associated with selective carrier excitation in one of the two non-equivalent K valleys (K$_+$ or K$_-$, respectively)~\cite{PRB11zz, PRL13xl, NComm12tc, PRB12gs, APL12gk, Nnt12km, Nnt12hz}. This spin-valley locking effect requires that a spin relaxation of a carrier is simultaneously accompanied by a valley change (i.e., an intervalley spin relaxation). As a consequence very long electron or hole spin/valley relaxation times -- in the microsecond range -- were measured in various TMD monolayers~\cite{Nphy15ly, PRL17pd, ACSNL19me,yalcin2024spinrelaxationlocalizedelectrons}. For the exciton (2-particles electron-hole system), numerous works have also shown that spin relaxation is governed by a very efficient mechanism linked to the electron-hole exchange interaction, leading to much shorter depolarization times, of the order of a few ps~\cite{PRB14cz, PRB14ty, Pssb15nk}. 
On the other hand, intra-valley spin relaxation mechanisms -- well known in III-V, II-VI and IV semiconductors~\cite{OO84fm, SSQC02da, SPS08md, PR10mw} -- are very poorly understood in TMDs. The intra-valley spin relaxation was mainly theoretically investigated  in \mos2 monolayer for in-plane carrier spin components~\cite{PRB14lw, PRB13ho}.

\begin{figure*}[tb]
\centering
\includegraphics[width=0.99\textwidth]{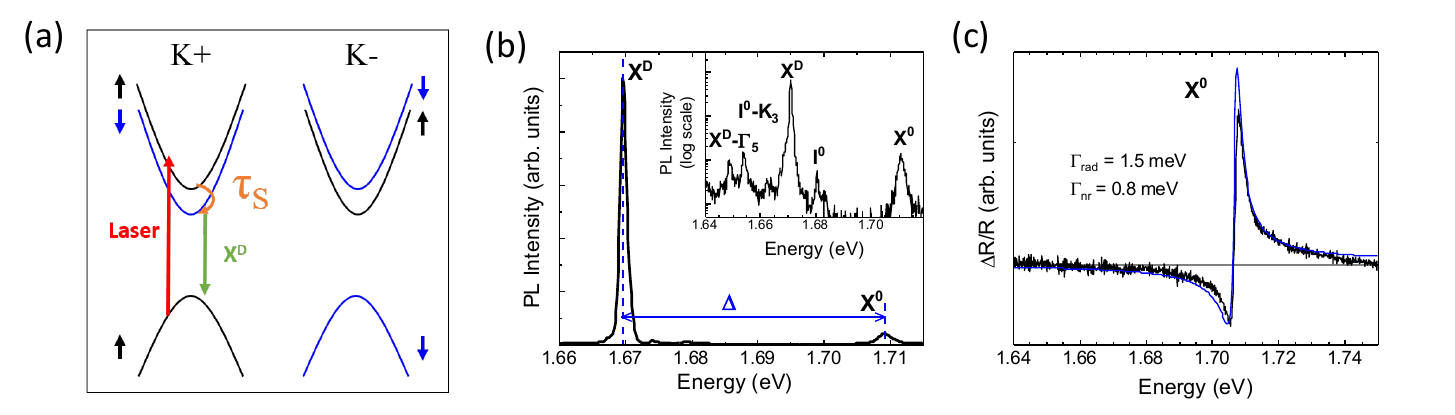}
\caption{(a) Single particle band structure of a \wse2 monolayer, showing the two inequivalent valleys K$_+$ and K$_-$ and the intra-valley electron spin relaxation from top to bottom conduction band, characterized by the time $\tau_s$; (b) Photoluminescence spectrum following a He-Ne laser excitation (1.96 eV); the splitting between the bright $X^0$ and the dark $X^D$ exciton is $\Delta=40~$meV, the inset corresponds to a blow-up of the spectrum in log scale, showing the weak emission components corresponding to the zero-phonon emission of the indirect exciton $I^0$ and its $K_3$ phonon replica $I^0- K_3$, together with the $\Gamma_5$ phonon replica of the dark exciton $X^D-\Gamma_5$; (c) Differential reflectivity spectrum evidencing the strong resonance of the bright exciton $X^0$ and the corresponding fit with the radiative and non-radiative broadening, $\Gamma_{rad}$  and $\Gamma_{nr}$ respectively (see text).}
\label{fig:fig1}
\end{figure*}

In this work we  investigate the electron and exciton spin properties of \wse2 monolayers. We demonstrate experimentally and theoretically that the intra-valley electron spin relaxation plays an important role in the exciton dynamics. Combining  time-resolved photoluminescence (TRPL) and stationary photoluminescence excitation (PLE) measurements, we show that a significant (spin-forbidden) dark neutral exciton emission occurs as a consequence of an efficient intra-valley electron spin relaxation yielding bright to dark exciton transfer. In a simplified single particle picture, the laser excitation generates bright excitons $X^0$ composed of electron-hole pairs with the same spins from the top valence band and the upper conduction band in a given valley (K$_+$ or K$_-$), see Fig.~\ref{fig:fig1}(a). The lower in energy dark exciton $X^D$ involves electron from the bottom conduction band and hole from the top valence bands with anti-parallel spins~\cite{PRL15xz, NComm15gw, Nnt17xz, PRL19mm}.
Surprisingly, but in agreement with previous works~\cite{PRL17gw, Ncomm20mh} the PL spectra of undoped \wse2 monolayers are dominated by dark exciton $X^D$ recombination, see Fig.~\ref{fig:fig2}(b). This means that there is an efficient intra-valley electron spin relaxation process  (characterized by the time $\tau_s$) which yields the transfer of exciton from bright to dark states. This is the subject investigated in this paper with a two-fold purpose: (i) to identify the spin relaxation mechanism and (ii) to  determine its characteristic time $\tau_s$. Our experimental and theoretical analysis highlights the key role played by the  chiral phonons $\Gamma_5$ which induce this bright to dark exciton relaxation. These phonons were identified theoretically in the pioneering work by Song and Dery~\cite{PRL13ys}. From symmetry point of view, these phonon states transform as in-plane components of magnetic field $(B_x, B_y)$ and, hence, were called chiral phonons~\cite{Sci18hz, 2dm19hc}. Despite numerous studies of dark excitons, charged excitons or interlayer excitons,~\cite{Ncomm19lz, PRR19el, ACSnan19lz, 2dm18ms,PRL20le, Ncomm20mh, 2dm20ad, 2dm22sd} the role of these chiral phonons on intra-valley electron spin relaxation was not demonstrated to the best of our knowledge.

The organization of the paper is as follows. The sample fabrication and the stationary and time-resolved optical spectroscopy set-ups are described in Sec.~\ref{sec:Exptdet}. The experimental results  are presented in Sec.~\ref{sec:Exper}, and Sec.~\ref{sec:theory} details the theoretical interpretation and the modelling of the excitation of photoluminescence spectra. Conclusion is given in Sec.~\ref{sec:Concl}.

\section{Sample fabrication and optical spectroscopy set-ups}\label{sec:Exptdet}
We have fabricated a high quality van der Waals heterostructure  made of an exfoliated \wse2 monolayer embedded in  hBN crystals using a dry stamping technique~\cite{2dm14acg, PRX17fc}. First the \wse2 ML flake is prepared by micro-mechanical cleavage of a bulk crystal (from 2D Semiconductors) and deposited on a bottom layer of hexagonal boron nitride on SiO$_2$/Si substrates. Subsequently, hBN was deposited on top of the \wse2 ML. Continuous-wave (cw) micro-photoluminescence and excitation of photoluminescence experiments are performed with a narrow-line single frequency tunable Ti:Sapphire ring laser (Matisse, Spectra Physics), linearly polarized, focused on a spot diameter of $\sim 1~\mu$m. The PL signal is dispersed by a spectrometer and detected by a charged-coupled device camera. The white light source for reflectivity measurements is a halogen lamp with a stabilized power supply. Time-resolved PL measurements were performed using a mode-locked Ti:sapphire oscillator delivering $\sim 1.5~$ps pulses with a repetition rate of $80~$MHz and a Hamamatsu synchro-scan streak camera with an overall instrument response time resolution of $\sim5~$ps (which includes both the  streak camera resolution and the dispersion induced by the diffraction grating). All the experiments are performed in a closed cycle cryostat at $T=4~$K.

\section{Experimental results}\label{sec:Exper}

\begin{figure*}[t]
\centering
\includegraphics[width=16.0 cm]{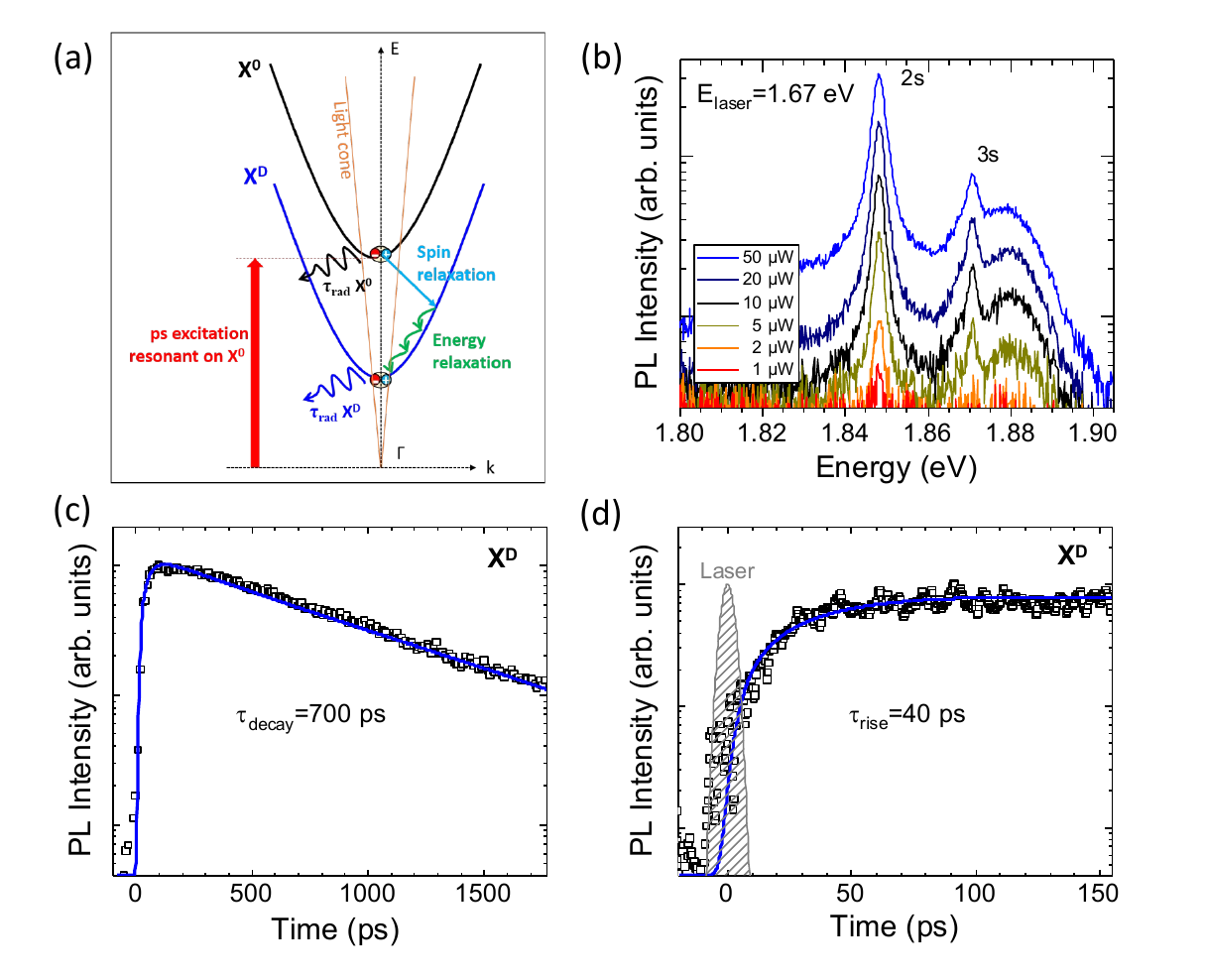}
\caption{(a) Schematics of the exciton dispersion curves with a simplified sketch of the transfer from bright to dark exciton branch where the 
excitation laser is strictly resonant with the bright exciton $X^0$ energy; (b) PL spectra for different average excitation laser power, showing the 
luminescence emission of 2s and 3s exciton excited state, with larger energy than the excitation laser ($E_{laser}=1.71~$ eV), resonant with the 
bright exciton 1s ground state; (c) Dark exciton $X^D$ photoluminescence dynamics following a ps laser excitation resonant with the 1s bright 
exciton $X^0$ ground state, the full dashed blue line is a bi-exponential fit with a decay time $\tau_{\rm decay}=700~$ ps and a rise time $\tau_{\rm rise}=40~$ ps,  laser power 1$~\mu$W; (d) Blow-up of the dark exciton $X^D$ PL rise time [cf. Eq.~\eqref{sol:ND:1}]}.  
\label{fig:fig2}
\end{figure*}

Figure~\ref{fig:fig1}(b) presents the photoluminescence (PL) spectrum of the \wse2 monolayer following cw He-Ne laser excitation (1.96 eV). 
In addition to the recombination of the bright exciton $X^0$, we observe a strong signal associated to the emission of the dark exciton $X^D$, with a 
bright-dark splitting of $\Delta\approx40~$meV in agreement with previous reports~\cite{PRL17gw, Ncomm20mh, CP21pk}. Its radiative 
recombination for in-plane polarized light is forbidden. However, it is allowed for $z$-polarized light~\cite{glazov2014exciton,PRB16je,PRL17gw}; 
the $z$-axis is perpendicular to the monolayer plane. The clear observation of $X^D$ in the PL spectrum, as shown in Fig.~\ref{fig:fig1}(b), is the 
consequence of using high numerical aperture objective (NA$=0.82$) which allows us to detect the $z$-polarized luminescence component. This 
strong $X^D$ emission, observed in many works, means that there is an efficient relaxation process from bright to dark exciton, which implies a 
valley conserving electron spin relaxation. We note that the spectrum of Fig.~\ref{fig:fig1}(b) is dominated by both bright and dark neutral excitons 
with negligible emission of trions. This shows that the sample is characterized by low doping densities and as a consequence the possible bright-
dark relaxation processes induced by free carriers are very unlikely~\cite{PRB22my}. Figure~\ref{fig:fig1}(c) presents the measured differential 
reflectivity spectrum, together with a fit using the transfer matrix method~\cite{PRM18cr}. The data are well described with $\Gamma_{rad}
=1.5~$meV and  $\Gamma_{nr}=0.8~$meV corresponding to the radiative and non-radiative broadening of the neutral exciton $X^0$, respectively. 
In the calculation we used the following values of  the refractive index $n_{\rm hBN}=2.2$, $n_{\rm SiO_2}=1.46$, $n_{\rm Si}=3.5$ and the different 
layer thicknesses of the top hBN$~=15~$nm, bottom hBN $~=113~$nm (measured by atomic force microscopy)\, and 
SiO$_2=83~$nm~\cite{PRL19hf, PRL23lr}.

To get more information on the relaxation dynamics, we have performed TRPL measurements with a laser excitation energy strictly resonant with the bright exciton $X^0$ energy, see the schematics in Fig.~\ref{fig:fig2}(a). We took special care to check that these experiments were performed with low enough laser excitation densities in order to avoid Auger type effects, which could affect the relaxation dynamics~\cite{PRX18bh}. Indeed, the resonant photo-generation of a significant density of bright excitons can yield an efficient exciton up-conversion effect due to exciton-exciton annihilation. This process leads to the generation of high energy exciton states (above the $X^0$ ground state), as attested by the detection of luminescence components with higher energy than the laser excitation one~\cite{Manca:2017aa,PRX18bh,Lin:2022aa}. Figure~\ref{fig:fig2}(b) illustrates this upconversion effect when the ps excitation laser is resonant with the $X^0$ ground state:  for laser powers larger than 1$~\mu$W, significant PL intensities corresponding to the recombination of 2s and 3s excited states, above the $X^0$ 1s ground state are clearly observed~\cite{PRL14ac}.

Hence, for all the experiments presented below, we have tuned the laser excitation power so that the Auger effects are negligible. Figure~\ref{fig:fig2}(c) displays the time evolution of the dark exciton $X^D$ PL intensity in these conditions after a ps excitation pulse. A bi-exponential fit yields a rise time $\tau_{\rm rise}= 40~$ps and a decay time $\tau_{\rm decay}=700~$ps. The latter corresponds to the dark exciton lifetime, already measured by different groups~\cite{Ncomm19yt, PRL23lr}. Remarkably, this measured dark exciton dynamics implies that the electron intra-valley spin relaxation from the upper to lower conduction band is efficient. However the calculated dark exciton dynamics detailed in Sec.~\ref{subsec:kin} shows that the rise time $\tau_{rise}$ does not correspond to the electron spin relaxation time $\tau_s$; the PL rise time includes in addition the energy relaxation time within the dark exciton branch $\tau_\epsilon$ that is on the order of 10's ps [see Fig.~\ref{fig:fig2}(a)] ~\cite{PhysRevLett.124.166802,Rosati:2020aa}. 

\begin{figure}[h]
\centering
\includegraphics[width=\linewidth]{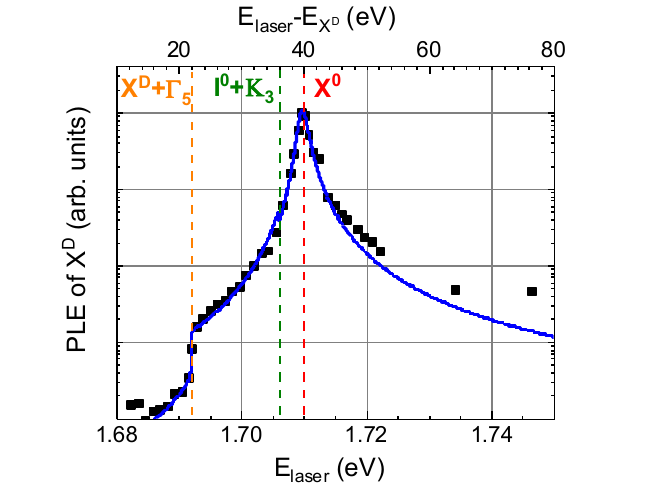}
\caption{Photoluminescence excitation spectrum with a detection energy corresponding to the dark exciton $X^D$ emission (1.67 eV); the bottom and top axis corresponds to the laser energy and the energy difference between the laser and the dark exciton energy ($E_{laser}-E_{X^D}$) respectively.The full blue line is the result of the fit and the model presented in Sec.~\ref{sec:theory}. The parameters are given in Tab.~\ref{tab:params}}.
\label{fig:fig3}
\end{figure}

To better understand the relaxation mechanism, we have measured with great accuracy the dark exciton $X^D$ PL as a function of the cw excitation laser energy. The corresponding PLE spectrum  is displayed in Fig.~\ref{fig:fig3} (note the dynamic range of four orders of magnitude). As expected, the strongest signal corresponds to the laser energy strictly resonant with the bright exciton $X^0$ energy: The corresponding strong absorption is followed by an efficient electron spin-flip and further energy relaxation down to the dark exciton, Fig.~\ref{fig:fig2}(a). Interestingly we observe a very clear jump of the PLE signal for $E_{\rm laser}-E_{X^D} =22~$meV. This energy corresponds exactly to the energy of the chiral phonon $\Gamma_5$ ($\hbar\Omega_5=22~$meV), demonstrating its key role on the intra-valley electron spin relaxation inducing the bright to dark exciton transfer~\cite{PRL13ys, PRB14zj, Ncomm20mh}. Note that under the conditions
\begin{equation}
\label{conds}
\hbar\Omega_5 < E_{\rm laser}-E_{X^D} < \Delta
\end{equation}
the bright state is a virtual state since the difference of energy between the laser and the dark exciton is smaller than the bright-dark splitting, but the final state in the process $X^D$ is the real state.  For $E_{laser}-E_{X^D} < \hbar\Omega_5$, the dark exciton luminescence is vanishingly small since there is no efficient process which allows the electron  relaxation from the top to the bottom conduction band with opposite spins. In addition, Fig.~\ref{fig:fig3}(a) shows a slight change of the PLE slope for  $E_{\rm laser}-E_{X^D} =36~$ meV; remarkably this value corresponds to the energy of $I^0$+$K_3$ state, where $I^0$ is the indirect exciton energy and $\hbar\Omega_3= 26$~meV the energy of the inter-valley $K_3$ phonon. Note that the energy $I^0$ can be determined thanks to the detection of the weak defect-induced (zero-phonon) assisted transition, see the inset in Fig. 1(b)~\cite{Ncomm20mh, PRB22pl}. This suggests that the scattering from direct bright, $X^0$ to indirect, $I^0$, excitons is also important. 

\section{Theory and Modelling}\label{sec:theory}

This section presents the theory and outlines the model of the PLE spectrum in Fig.~\ref{fig:fig3}. We start with an analysis of the selection rules for the intra-valley spin flip process (Sec.~\ref{subsec:sel}), continue with the model of the PLE (Sec.~\ref{subsec:PLE}), describe the specifics of the spectra in van der Waals heterostructures related to the multiple reflections of light from the interfaces and fit the experimental data in Sec.~\ref{subsec:fit}. We briefly discuss the kinetics of the dark exciton emission in Sec.~\ref{subsec:kin}.

\subsection{Electron spin-flip process and bright-dark exciton relaxation}\label{subsec:sel}
Let us first analyze from the symmetry standpoint the microscopic mechanism of the intra-valley electron spin relaxation which leads to the bright-dark exciton relaxation and present its qualitative picture. In the simple single particle band structure displayed in Fig.~\ref{fig:fig1}(a), it is usually assumed that the top (bottom) conduction band is described by pure spin $\uparrow$ ($\downarrow$) state, respectively, for the K$_+$ valley. This is an oversimplified picture: It was shown that the spin-orbit induced coupling with higher energy bands yields a spin mixing. Within a simple model accounting for the lowest orbital conduction band (even at the monolayer plane reflection) and first excited conduction band (odd at the mirror reflection), the eigenstate for the bottom CB writes~\cite{PRB16je, PRL19hf}:
\begin{subequations}
\label{so-mixing}
\begin{equation}
\label{downarrow}
 |CB\downarrow\rangle=\beta \frac{(\mathcal X+\rm{i}\mathcal Y)}{\sqrt{2}}\downarrow+~\alpha \mathcal Z\uparrow ,
\end{equation}
where $(\mathcal X+\rm{i}\mathcal Y)/\sqrt{2}$ is the orbital Bloch function of the bottom conduction band at the K$_+$ valley, $\mathcal Z$ is the orbital Bloch function of the excited conduction band in the same valley, $\uparrow$ and $\downarrow$ describe the basic spinors, and $\alpha$ and $\beta$ are coefficients with $|\alpha|^2+|\beta|^2=1$. At the same time, for the top conduction band the eigenfunction reads
\begin{equation}
\label{uparrow}
 |CB\uparrow\rangle= \frac{(\mathcal X+\rm{i}\mathcal Y)}{\sqrt{2}}\uparrow,
\end{equation} 
with no admixture of the $\mathcal Z\downarrow$ state. Such an asymmetry follows from the fact that the states $ |CB\downarrow\rangle$ and $|CB\uparrow\rangle$ are not related by the time-reversal symmetry: Their Kramers-conjugate (time-reversal-related) counterparts belong to the opposite valley K$_-$. 

\end{subequations}

Density functional theory (DFT) calculations performed at the GW level show that $|\alpha|^2\approx0.02$ and $\beta\approx1$, with a weak dependence as a function of the wavevector within the relevant exciton dispersion range~\cite{PRB16je}. This value is consistent with estimates based on the tight-binding model presented in Ref.~\cite{PRB13kk}. Thanks to the non-zero $\alpha$ value, electrons in the top conduction band with majority spin $\uparrow$ could relax to the bottom CB with majority spin $\downarrow$ following a spin-conserving phonon scattering. The corresponding spin-flip process is close to the well-known Elliott-Yafet mechanism~\cite{OO84fm}. However, note that in contrast to other semiconductors such as GaAs, there is no Van Vleck cancellation here, since the admixture for the top and bottom CB is different~\cite{ENSD18mg}, compare the wavefunctions in Eqs.~\eqref{so-mixing}. This is because the top and bottom conduction band in a given valley are not related by the time reversal symmetry, as mentionned above the latter symmetry operation involves the change of the valley. It makes intra-valley spin relaxation in TMD semiconductors very unique.

Now we discuss the phonon that allows the transition between the state~\eqref{downarrow} and \eqref{uparrow}. From symmetry considerations,
 it turns out that this scattering is allowed for the chiral $\Gamma_5$ phonon~\cite{PRL13ys}: Indeed, the initial and final states differ in angular momentum component by $1$, hence, the phonon should transform as in-plane components of a magnetic field. Assuming that electron-phonon interaction conserves 
 spin (we disregard the spin-orbit contributions in the electron-phonon interaction matrix elements), the $\Gamma_5$ assisted phonon scattering 
 time from the top to the bottom CB takes the form:
\begin{equation}
\label{electron:taus}
\frac{1}{\tau_s}=\frac{2\pi}{\hbar}\sum_{k'}|M_{\Gamma_5}|^2\delta[\epsilon_{K\downarrow}(\bm k)-\epsilon_{K\uparrow}(\bm k')-\hbar\Omega_{5}]=\frac{\alpha^2}{\tau_{\Gamma_5}}.
\end{equation}
Here $\epsilon_{K\uparrow,\downarrow}(\bm k)$ are the corresponding electron dispersions with $\bm k$ being the electron wavevector in the initial state and the summation is carried out over all possible momenta of the final state, $\tau_{\Gamma_5}$ the characteristic time of phonon emission:
\begin{equation}
\frac{1}{\tau_{\Gamma_5}}=\frac{m}{2\pi\hbar^2}|D_{\Gamma_5}|^2    
\end{equation}
where $m$ is the electron effective mass and $D_{\Gamma_5}$ the corresponding electron-phonon coupling constant: $M_{\Gamma_5}=\alpha D_{\Gamma_5}$. According to the estimates of the matrix elements presented in Sec.~\ref{subsec:fit} $\tau_s \sim 10$~ps.

\subsection{Calculation of the PLE spectra}\label{subsec:PLE}

We consider the photoluminescence excitation effect where the emission of the dark exciton $X^D$ is detected following laser excitation with energy $E> E_{X^D}$. We consider the following formation path of the dark exciton: virtual creation of the bright exciton $X^0$ (with zero in-plane wavevector) subsequent emission of the $\Gamma_5$ phonon with energy $\hbar\Omega_5$ (its dispersion is neglected) followed by relaxation of $X^D$ towards its ground state where it emits a photon. Theoretically, the complexity of the problem is related to the fact that in the relevant energy range the bright exciton is in the virtual state and the kinetic equation is not directly applicable.

In the time-integrated PLE, the observed signal is proportional to the rate of the dark exciton generation rate $W_D$. Namely,
\begin{equation}
\label{PLE:Dark}
I_{\rm PLE} = P W_D,
\end{equation}
where the factor $P$ accounts for the interplay of the radiative and non-radiative recombination channels as well as the energy relaxation. The rate $W_D$ can be written as~\cite{PRB18ss} 
\begin{equation}
\label{dark:rate}
W_{D} = \frac{2\pi}{\hbar} \sum_{\bm k} |M^{(2)}|^2 \delta(E - E_{X^D_{\bm k}}-\hbar\Omega_5),
\end{equation}
where $E\equiv E_{\rm laser}$ is the laser energy $\bm k$ is the dark exciton in-plane wavevector, $E_{X^D_{\bm k}}$ is the dark exciton dispersion, and $M^{(2)}$ is the corresponding matrix element that should be calculated taking into account both exciton-photon and exciton-phonon interaction.

For $E_{X^0}-E\gg \Gamma$, where $E_{X^0}$ is the bright exciton energy, $E \equiv \hbar\omega$ is the incident photon energy, and  $\Gamma$ is the broadening of the $X^0$ state, $M^{(2)}$ can be evaluated with the second-order perturbation theory yielding,
\begin{equation}
\label{M2}
M^{(2)} = \frac{M_{\Gamma_5} M_{X^0}^{opt}}{E - E_{X^0}}.
\end{equation}
Here $M_{X^0}^{opt}$ is the matrix element of the transition with formation of the bright exciton, $M_{\Gamma_5}$ is the matrix element of the phonon-assisted conversion $X^0\to X^D$. Strictly speaking, $M_{\Gamma_5}$ for excitons in Eq.~\eqref{M2} differs from that for electrons, Eq.~\eqref{electron:taus} by a form-factor depending on the shape of exciton envelope functions. For  relevant wavevectors the form-factor is close to unity and omitted hereafter. Since the phonon-induced transition is allowed at $\bm k=0$ we can neglect the wavevector dependence  of the matrix element for simplicity. The summation over $\bm k$ gives the dark exciton density of states which is a constant function of energy in the parabolic approximation for the exciton dispersion. As a result, the PLE intensity is given by

\begin{equation}
  \label{PLE}
I_{\rm PLE} = P W_D  
= \frac{2P}{\hbar} \frac{|M_{X^0}^{opt}|^2}{(E-E_{X^0})^2}\gamma(E),  
\end{equation}
where 
\begin{align}
\label{gamma:5}
\gamma(E) = \gamma_5 \theta(E - E_{X_D} - \hbar\Omega_5), \\ 
\gamma_{5} 
= \pi \mathcal D_{D} |M_{\Gamma_5}|^2,\nonumber
\end{align} 
is the scattering rate related to the $\Gamma_5$ phonon emission. Here $\theta(x)=1$ for $x>0$ and $0$ for $x<0$ is the Heaviside $\theta$-function, $m_{X^D}$ is the translational mass of the dark exciton, hence, $E_{X_D,\bm k} = E_{X^D} + \hbar^2 k^2/(2m_{X^D})$, and $\mathcal D_D = m_{X^D}/(2\pi  \hbar^2)$ is the density of states of the dark exciton.  The dependence given by Eq.~\eqref{PLE} corresponds to the onset of the absorption tail of the $X^0$ state. 

Equation~\eqref{PLE} describes the simplest possible process with a single phonon scattering. Naturally, it is valid for large detunings only. Following Refs.~\cite{PRB18ss, Ncomm21ip, PRL17dc} we extend this expression to allow for arbitrary detunings taking into account other scattering contributions in addition to $\Gamma_5$ phonons. First, we consider the ML alone; the multiple layers corresponding to the real sample will be considered in Sec.~\ref{subsec:fit}. The self-energy of the bright exciton outside the light-cone can be written in the form
\begin{equation}
\label{exciton:Sigma}
\Sigma(E) = - \mathrm i \left[\gamma(E) + \delta \gamma(E)\right],
\end{equation}
where  $\delta \gamma(E)$ takes into account all remaining relaxation channels, see below. In Eq.~\eqref{exciton:Sigma} we consider only the imaginary part of the self-energy, its real part is included in the exciton transition energy. The $\delta\gamma(E)$ can be expressed as a sum of two contributions
\begin{equation}
\label{delta:Gamma:gen}
\delta \gamma(E) = \gamma'(E) + \gamma_e(E),
\end{equation}
which include\\
\indent 1. The conversion of bright excitons into indirect intervalley exciton $I^0$ 
\begin{equation}
\label{gamma'}
\gamma'(E) = \gamma_3 \theta(E - E_{X'} - \hbar\Omega_3) + \gamma_2 \theta(E - E_{X'} - \hbar\Omega_2),
\end{equation} 
with the symmetry allowed  ($K_3$) and symmetry forbidden ($K_2$) phonons~\cite{PRL13ys, PRB19mg, Ncomm20mh} ($E_{I^0}$ is the indirect exciton energy, $\hbar\Omega_{2,3}$ are the corresponding phonon energies), and\\
\indent 2. The elastic or quasielastic scattering of bright excitons by static defects and acoustic phonons
\begin{equation}
\label{gamma:e}
\gamma_e(E) = \gamma_{sc} \theta(E - E_{X^0}).
\end{equation}
Hereafter we consider low temperatures where the phonon absorption processes can be disregarded. For the same reason, we consider only spontaneous phonon emission in Eqs.~\eqref{electron:taus} and \eqref{M2}.

Typical scattering pathway includes arbitrary number of elastic or quasi-elastic scatterings of the photoexcited bright exciton and emission of the single $\Gamma_5$ phonon that transfers the bright exciton to a dark branch. The additional scattering between the indirect and dark excitons is neglected here for simplicity. As a result, 
\begin{multline}
\label{dark:rate:full}
W_{D} = \frac{2\pi}{\hbar} \sum_{\bm k} \frac{|M_{\Gamma_5}M_{X^0}^{opt}|^2}{(E - E_{X^0})^2 + \left[\gamma(E) + \delta \gamma(E) + \Gamma_{rad}\right]^2}
\\ 
\frac{\gamma(E) + \delta \gamma(E)}{\gamma(E) + \delta \gamma(E) - \gamma_e(E)}
\delta(E - E_{X^D_{\bm k}}-\hbar\Omega)
\\
= \frac{2}{\hbar} 
\frac{|M_{X^0}^{opt}|^2}{(E - E_{X^0})^2 + \left[\gamma(E) + \delta \gamma(E)+ 
\Gamma_{rad}\right]^2}
\\
\frac{\gamma(E) + \delta \gamma(E)}{\gamma(E) + \delta \gamma(E) - \gamma_e(E)} 
\gamma(E).
\end{multline}
Here, as before, $\Gamma_{rad}$ is the exciton radiative decay rate and the factor 
\begin{equation}
\frac{\gamma(E) + \delta \gamma(E)}
{\gamma(E) + \delta \gamma(E) - \gamma_e(E)} 
= \frac{1}{1- \frac{\gamma_e(E)}{\gamma(E) + \delta \gamma(E)}}
\end{equation}
accounts for multiple elastic (or quasielastic) scattering processes (we assume that the probability of an intermediate state to be in the light cone is negligible). Finally, the PLE intensity can be presented as: 
\begin{multline}
\label{IPLE:tot}
I_{\rm PLE} = \frac{2P}{\hbar} \frac{\gamma(E)}{\gamma(E) + \gamma'(E) }\\
\times  \frac{|M_{X^0}^{opt}|^2[\gamma(E) + \delta \gamma(E)]}{(E - E_{X^0})^2 + \left[\gamma(E) + \delta \gamma(E) + \Gamma_{rad} \right]^2}
 .
\end{multline}
Equation~\eqref{IPLE:tot} can be transformed to another form that highlights its physical meaning by introducing the absorption coefficient of the monolayer as 
\begin{equation}
\label{absorption:gen}
\mathcal A(E) = \frac{2\Gamma_{rad} [\gamma(E)+ \delta \gamma(E)]}{(E-E_{X^0})^2 + [\Gamma_{rad}+ \gamma(E)+ \delta \gamma(E)]^2}.
\end{equation} 
The PLE intensity can be written in the following simple form:
\begin{equation}
\label{IPLE:tot:1}
I_{\rm PLE} = \frac{I}{E} \times P \times \mathcal A(E) \frac{\gamma(E)}{\gamma(E) + \gamma'(E)}.
\end{equation} 
Here $I$ is the incident laser intensity ($I/E$ is the flux density of incident photons). In Eq.~\eqref{IPLE:tot:1} the factor $\mathcal A(E)$ gives the probability for the incident photon to be absorbed and the factor $\gamma(E)/[\gamma(E) + \gamma'(E)]$ accounts for the fact that only excitons scattered by the $\Gamma_5$ phonons arrive at the dark state and emit light.

\subsection{PLE spectra of the van der Waals heterostructure}\label{subsec:fit}

In a van der Waals heterostructure the situation is more involved owing to the presence of multiple reflections of light from the interfaces. As a result, absorption spectra (and, correspondingly, the photoluminescence excitation spectra) differ from simple Eq.~\eqref{absorption:gen}. To take these effects into account we, first, evaluate the exciton self-energy due to the coupling with $\gamma_5$ phonons and other scattering mechanisms as Eqs.~\eqref{exciton:Sigma} -- \eqref{gamma:e}. We use it to determine the monolayer reflectivity:
\begin{multline}  
\label{r:ML}
r_{ML} = - \frac{\mathrm i \Gamma_{rad}}{E - E_{X_0} - \Sigma(E) + \mathrm i \Gamma_{rad}}  \\
= - \frac{\mathrm i \Gamma_{rad}}{E - E_{X_0} + \mathrm i [\Gamma_{rad} + \gamma(E)+ \delta \gamma(E)]}. 
\end{multline}
This reflection coefficient together with the transmission coefficient $t_{ML}=1+r_{ML}$ is used to construct the monolayer transfer matrix which, in turn, is substituted to the transfer matrix of the van der Waals heterostructure: ``top hBN/ML/bottom hBN/SiO$_2$/Si'' following Refs.~\cite{PRM18cr,PRL19hf,PRL23lr}. This transfer matrix allows us to calculate the absorption $\mathcal A_{\rm vdW}(E)$ and reflection spectra of the structure. In actual fit the $\theta$-function in Eq.~\eqref{gamma:e} was slightly broadened replacing $\theta(x)$ by $1/2+\arctan[x/(\Gamma_{0}+\Gamma_{nr})]$ to avoid unphysical spike in the reflectivity. The photoluminescence excitation spectra is given by analogy with Eq.~\eqref{IPLE:tot:1} 
\begin{equation}
\label{IPLE:tot:vdW}
I_{\rm PLE, vdW} = \frac{I}{E} \times P \times \mathcal A_{\rm vdW}(E) \frac{\gamma(E)}{\gamma(E) + \gamma'(E)}.
\end{equation}
Equation~\eqref{IPLE:tot:vdW} is used to plot the blue curve in Fig.~\ref{fig:fig3}.

\begin{table}[b]
\caption{Parameters used in the modelling of Figs.~\ref{fig:fig3} and \ref{fig:fig4}. \label{tab:params}}
\begin{ruledtabular}
\begin{tabular}{c|cc}
Parameter & Value \\
\hline
Bright exciton energy $E_{X_0}$ & $1.7100$~eV\\
Dark exciton energy $E_{X^D}$ & $1.6700$~eV\\
Indirect exciton energy $E_{X'}$ & $1.6800$~eV\\  
$\Gamma_2$ phonon energy $\hbar\Omega_2$ & $13$~meV\\
$\Gamma_3$ phonon energy $\hbar\Omega_3$ & $26$~meV\\
$\Gamma_5$ phonon energy $\hbar\Omega_3$ & $22$~meV\\
$\Gamma_2$ phonon scattering rate $\gamma_2$ & $0$~meV\\
$\Gamma_3$ phonon scattering rate $\gamma_2$ & $0.11$~meV\\
$\Gamma_5$ phonon scattering rate $\gamma_2$ & $0.08$~meV\\
Quasielastic scattering rate $\gamma_{sc}$ & $0.61$~meV\\
Non-radiative decay rate $\Gamma_{nr}$ & $0.8$~meV\\
Radiative decay rate $\Gamma_{rad}$ & $1.5$~meV
\end{tabular} 
\end{ruledtabular}
\end{table}

As many parameters are involved, we use the following procedure to fit the PLE spectrum (see Tab.~\ref{tab:params} for the complete list of parameters):\\
\indent 1. The radiative and non-radiative broadening of the bright exciton are extracted from the measured reflectivity spectrum (see Fig.~\ref{fig:fig1}(c)): $\Gamma_{rad} =1.5~$meV and $\Gamma_{nr} =0.8~$meV.\\
\indent 2. We use the calculated broadening $\gamma_3$ induced by the intervalley phonon $K_3$:\\
\begin{equation}
\label{gamma:3:rate:est}
\gamma_3 = \pi \mathcal D_{D}  |M_{K_3}|^2  ,
\end{equation}
where we assumed that the density of states of the intervalley exciton and of the dark exciton is the same. We use the following expressions~\cite{PRB19mg}
\begin{equation}
M_{K_3} = \sqrt{\frac{\hbar}{2\rho \Omega_3}}D_3,
\end{equation}
where $D_3 \approx 10^{8}$~eV/cm is the corresponding interaction constant [cf. Ref.~\cite{PRB14zj}], $\rho=6\times 10^{-7}$~gm/cm$^2$ is the mass density~\cite{PRB18ss}, $\hbar\Omega_3= 26$~meV is the $K_3$ phonon energy, and $m_D = 0.76m_0$ is the dark exciton effective mass. This yields $\gamma_3 =0.11~\mbox{meV}$. Note that this estimate is somewhat smaller than the one reported in Ref.~\cite{Ns19sb}. However, the calculated phonon-induced broadening in Ref.~\cite{Ns19sb} would yield a PL linewidth 3 times larger than the experimental one measured in Fig.~\ref{fig:fig1}(c). \\
\indent 3. We neglected the broadening due to $K_2$ symmetry-forbidden phonons: $\gamma_2=0$. Indeed the measurements show that the $K_2$ phonon replica of exciton complexes are much weaker compared to the ones involving $K_3$ phonons~\cite{Ncomm20mh, PRB22my}.\\
\indent 4. Finally, we fit the dark exciton PLE spectrum to find the optimal value of the $\Gamma_5$ phonon scattering rate using $\gamma_{sc}$ in Eq.~\eqref{gamma:e} such that the total non-radiative broadening is fixed to be $\Gamma_{nr}$ extracted from the experiment: 
\[
\Gamma_{nr} = \gamma_{sc} + \gamma_2+\gamma_3 + \gamma_5.
\]
We find $\gamma_5=0.08$~meV and $\gamma_{sc}=0.61$~meV, see the full line in Fig.~\ref{fig:fig3}. This corresponds to a spin-relaxation time due to the $\Gamma_5$ chiral phonon of $\tau_s \sim 8$~ps. The corresponding matrix element for the exciton $\Gamma_5$ phonon interaction can be then evaluated as
\begin{equation}
\label{M:Gamma:5}
D_5 = D_3 \times \sqrt{\frac{\Omega_5}{\Omega_3}} \times \sqrt{\frac{\gamma_5}{\gamma_3}} \approx 9.6 \times 10^7~\mbox{eV/cm},
\end{equation}
giving a value close to the calculated $D_3$ value. Taking into account the fraction $|\alpha|^2=2\%$ of the $\mathcal Z\uparrow$ state admixture in the bottom conduction band, we can evaluate the spin conserving matrix element between the $\mathcal Z\uparrow$ and $(\mathcal X+ \mathcal Y)\uparrow/\sqrt{2}$ conduction band states as $D_5^{zxy} \approx 6.8\times 10^{8}$~eV/cm which is in the range of calculated values~\cite{PRB14zj, PRB12kk}. Interestingly, the spin relaxation time $\tau_s \sim 8$~ps we find here is perfectly consistent with the measured decay time of the negative triplet trion ( also governed by the intra-valley spin relaxation process)~\cite{PRB22my}. 

\begin{figure}[t]
\centering
\includegraphics[width=0.9\linewidth]{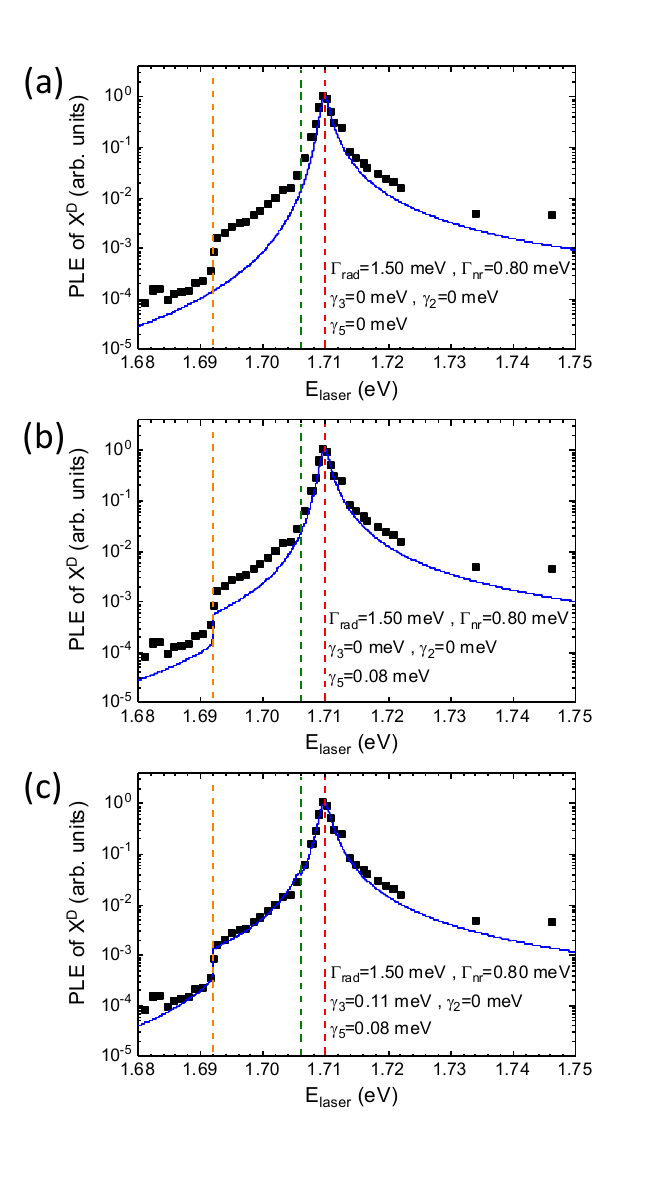}
\caption{Calculated photoluminescence excitation spectrum detected at the dark exciton $X^D$ for different values of phonon scattering rates: (a) $\gamma_3 = \gamma_5 =0$, (b) $\gamma_3 = 0$ and $\gamma_5 = 0.08$~meV, (c) $\gamma_3 = 0.11$~meV and $\gamma_5 = 0.08$~meV. All remaining parameters are fixed, see text and Tab.~\ref{tab:params}. Points show experimental data.}.
\label{fig:fig4}
\end{figure}

To demonstrate the respective contribution of the different parameters to the PLE spectra we show in Fig.~\ref{fig:fig4} the calculated PLE for three cases: $\gamma_3 = \gamma_5 =0$ [panel (a)], $\gamma_3 = 0$, $\gamma_5 = 0.08$~meV [panel (b)], and $\gamma_3 = 0.11$~meV, $\gamma_5 = 0.08$~meV [panel (c)]. One can clearly see that inclusion of both $\Gamma_3$ and $\Gamma_5$ phonon scattering processes allows to reproduce the experimental data.

\subsection{Calculation of the dark exciton PL dynamics}\label{subsec:kin}

To provide a physical picture of the TRPL dynamics measured in Fig.~\ref{fig:fig2}(c), we analyze the exciton dynamics within the kinetic equation framework and consider the occupancies of the bright excitons with small in-plane momenta $N_B$ (created by light), dark excitons with large momenta (outside the light cone) $N_D'$, and radiative dark excitons $N_D$. They obey a set of equations in the form
\begin{subequations}
\label{kinetic:BDD}
\begin{align}
&\frac{d N_B}{dt} + 2(\Gamma_B + \gamma)N_B = G(t),\\
&\frac{d N_D'}{dt} + \frac{N_D'}{\tau_\epsilon} = 2\gamma N_B,\\
&\frac{d N_D}{dt} + \frac{N_D}{\tau_D} = \frac{N_D'}{\tau_\epsilon}.
\end{align}
\end{subequations}
Here $G(t)$ is the bright exciton generation rate, and the remaining notations are as above: $\gamma$ is the $\Gamma_5$-phonon assisted damping of the bright exciton (to the dark one) corresponding to the chiral phonon spin relaxation time, $\Gamma_B$ includes both the radiative decay and all other scattering processes of the bright exciton, $\tau_\epsilon$ is the energy relaxation time of the dark exciton and $\tau_D$ is the (long) decay time of dark exciton. We assume that $\tau_D^{-1} \ll \tau_\epsilon^{-1} \ll \Gamma_B + \gamma \ll \tau_p^{-1}$ where $\tau_p$ is the laser pulse duration. 

The solutions of the set of Eqs.~\eqref{kinetic:BDD} can be recast as [with $N_0$ being the initial density of bright excitons $N_0 = \int_{-\infty}^\infty G(t) dt$]
\begin{subequations}
\label{solutions:BDD}
\begin{align}
N_B (t) &= N_0 e^{-2(\Gamma_B + \gamma)t}, \label{sol:NB}\\
N_D' (t) &= 2\gamma \int_0^t \exp{\left(\frac{t'-t}{\tau_\epsilon} \right)} N_B(t') dt'\\
&= N_0\frac{2\gamma\tau_\epsilon }{2(\Gamma_B + \gamma)\tau_\epsilon -1}\left[ e^{-t/\tau_\epsilon} - e^{-2(\Gamma_B + \gamma)t}\right], \nonumber\\
N_D(t) &= \frac{1}{\tau_\epsilon} \int_0^t \exp{\left(\frac{t'-t}{\tau_D} \right)} N_D'(t') dt' .
\end{align}
\end{subequations}
As a result,
\begin{multline}
\label{sol:ND}
N_D(t)\approx N_0\frac{2\gamma\tau_\epsilon }{2(\Gamma_B + \gamma)\tau_\epsilon -1} \left[ e^{-t/\tau_D} - e^{-t/\tau_\epsilon}\right] \\
+ N_0\frac{\gamma/(\Gamma_B + \gamma) }{2(\Gamma_B + \gamma)\tau_\epsilon -1}[e^{-2(\Gamma_B + \gamma)t} - e^{-t/\tau_D}].
\end{multline}
In Eq.~\eqref{sol:NB} we neglected the fast laser rise on the timescale of $\tau_p^{-1}$. The population of dark excitons with large wavevectors $N_D'$ rises on the timescale $\sim (\Gamma_B + \gamma)^{-1}$. The dynamics of dark exciton population $N_D$ is more involved but generally it rises on the timescale $\sim \tau_\epsilon$ and decays on the timescale $\sim \tau_D$. Note that in our conditions $\tau_\epsilon \gg (\Gamma_B + \gamma)^{-1}$ the second term in Eq.~\eqref{sol:ND} is small as compared to the first term and as a consequence,
\begin{equation}
\label{sol:ND:1}
N_D(t)\approx N_0\frac{\gamma }{\Gamma_B + \gamma} \left[ e^{-t/\tau_D} - e^{-t/\tau_\epsilon}\right].
\end{equation}
so, the dynamics is mainly controlled by the energy relaxation time $\tau_\epsilon$ and the lifetime $\tau_D$ of the dark exciton. This is exactly the behavior of the dark exciton emission intensity presented in Fig.~\ref{fig:fig2}(c,d). This analysis confirms that it is hardly possible to get a quantitative estimate of the spin relaxation time induced by the chiral phonons from the measurement of the PL dynamics as the one presented in Fig.~\ref{fig:fig2}(c) because its rise is controlled by the energy relaxation time. 

\section{Conclusion}\label{sec:Concl}
By a combined experimental-theoretical study, we have shown that the spin relaxation of an electron in a TMD monolayer can occur efficiently without a change of valley. One of the important consequences is the fast bright-dark exciton transfer as a result of the relaxation from top to bottom conduction band states with opposite spins. We estimate the intra-valley electron spin relaxation time to be $\sim 10$~ps in \wse2 monolayer. Despite some similarities with the well known Elliot-Yafet spin relaxation mechanism, we demonstrate that this intra-valley spin flip process in TMD semiconductors is very unique.
These results, which go beyond the simple approach based on the spin-valley locking effect, show the importance of the effects related to spin-orbit mixing and to the relaxation induced by chiral phonons in atomically thin crystals.

\acknowledgements
This work was supported by the Agence Nationale de la Recherche under the program ESR/EquipEx+ (Grant No. ANR-21-ESRE- 0025), ANR projects ATOEMS and IXTASE,  and the France 2030 government investment plan managed by the French National Research Agency under grant reference PEPR SPIN -- SPINMAT ANR-22-EXSP-0007. 

%

\end{document}